\documentclass[%
 reprint,
 amsmath,amssymb,
 aps,
]{revtex4-1}

\usepackage{graphicx}
\usepackage{dcolumn}
\usepackage{bm}
\usepackage[colorlinks,linkcolor=blue,anchorcolor=blue,citecolor=blue,urlcolor=blue]{hyperref}

\begin{document}

\preprint{APS/123-QED}

\title{Preparing multiparticle entangled states of NV centers via adiabatic ground-state transitions}

\author{Yuan Zhou}
\author{Bo Li}
\author{Xiao-Xiao Li}
\author{Fu-Li Li}
\author{Peng-Bo Li}
 \email{lipengbo@mail.xjtu.edu.cn}
\affiliation{Shaanxi Province Key Laboratory of Quantum Information and Quantum Optoelectronic Devices,
Department of Applied Physics, Xi'an Jiaotong University, Xi'an 710049, China}

\date{\today}

\begin{abstract}

We propose an efficient method to generate multiparticle entangled states of NV centers in a spin mechanical system, where
the spins interact through a collective coupling of the Lipkin-Meshkov-Glick (LMG) type.  We show that, through adiabatic
transitions in the ground state of the LMG Hamiltonian, the Greenberger-Horne-Zeilinger (GHZ)-type or the W-type entangled
states of the  NV spins can be generated   with this hybrid system from an initial product state. Because of adiabaticity,
this scheme is robust against practical noise and
experimental imperfection, and may be useful for quantum information processing.

\end{abstract}

\maketitle


\section{introduction}

In recent years, much attention has been paid to the generation of  multiparticle entangled states with different systems,
which play a key role in
quantum computation, quantum networks, quantum teleportation, and quantum cryptography
\cite{PhysRevLett.80.2245, RevModPhys.74.197, RevModPhys.81.865, PhysRevLett.82.1971,James1998Quantum, PhysRevLett.86.5188, PhysRevLett.119.090401}.
Thus far, a plenty of  schemes for preparing multiparticle entangled states have been proposed, with a variety of setups
such as ion traps, cavity QED, spin-mechanics, etc
\cite{PhysRevLett.74.4091, PhysRevLett.87.230404, PhysRevLett.82.1835, PhysRevLett.87.137902, PhysRevLett.90.133601, PhysRevLett.100.040403,PhysRevLett.100.170504, PhysRevLett.111.180401, PhysRevLett.117.040501, PhysRevLett.119.183602,PhysRevB.77.014510,Li2017,PhysRevA.86.022329,PhysRevA.87.022320}.
Furthermore,
some of these schemes have been successfully implemented  in  experiment \cite{H2005Scalable, PhysRevLett.76.1796, Barontini1317, Neumann1326, C2000Experimental, Zhu2011Coherent, Johnson2017Ultrafast}.
Especially,  hybrid quantum systems are reliable and promising setups
for quantum information processing due to  their easy scalability and longer coherence times \cite{RevModPhys.85.623,Rabl2006Hybrid,PhysRevB.85.024537,PhysRevB.87.144516, PhysRevA.88.012329, Song2017Dissipation, Houck2016On,Buluta2012Natural,PhysicsTodayNori}.

Among all microscopic solid state systems, nitrogen-vacancy (NV) centers  in diamond
are particularly attractive
due to their excellent spin properties even at ambient conditions
\cite{Sungkun2012Coherent, Doherty2013The, Maze2011Properties, PhysRevLett.105.140502, PhysRevLett.107.220501, PhysRevLett.111.227602, PhysRevLett.116.143602, PhysRevLett.113.020503, Bar2013Solid, Maze2011Properties, PhysRevB.85.205203}.
Significant theoretical and experimental investigations have been carried out to realize
 quantum logical gates, quantum state manipulating,
and entangled state generation
\cite{PhysRevLett.112.116403, PhysRevLett.113.237601, PhysRevX.6.041060}.
However, it is still a challenge to generate multipartite entanglement
among  distant NV centers in hybrid quantum systems \cite{PhysRevLett.108.066803,PhysRevLett.108.143601,PhysRevA.83.022302, PhysRevB.94.205118}.

In principle, the precondition for manipulating or entangling   NV spins is to
acquire the strong coupling between  the NV spins and other quantum  data buses \cite{PhysRevLett.117.015502, PhysRevLett.110.156402,PhysRevApplied.10.024011, PhysRevLett.107.060502, PhysRevB.90.195112}.
Much work has been proposed by taking advantage of the strong magnetic coupling between  NV center ensembles and
superconducting  microwave cavities or  qubits \cite{PhysRevLett.105.140502, PhysRevLett.107.220501, PhysRevLett.105.210501, PhysRevLett.113.023603, PhysRevA.96.062333}.
In fact, the more attractive investigation is the strong magnetic
coupling between  nanomechanical resonators (NAMR) and  single NV centers or a few of distant NV centers \cite{NatPhysWrachtrup, Albrecht2013Coupling,PhysRevB.79.041302,PhysRevA.80.022335,PhysRevA.96.023827}.
Based on the spin-mechanical system,
several promising theoretical schemes have also been proposed to
prepare  entangled NV spins by utilizing the NAMR as a data bus \cite{PhysRevLett.117.015502, PhysRevB.94.205118, PhysRevA.80.022335, PhysRevA.96.023827, P2010A}.
Since these schemes  naturally
rely on the dynamical evolution of the hybrid spin mechanical system,
the target state is inevitably disturbed by  dissipations and ambient thermal noises.
Therefore, it  is appealing  to propose a high-efficiency and more feasible protocol
for preparing multiparticle entangled NV spins.

In this work, we propose an efficient scheme for generating  multiparticle entangled states of NV centers in a spin
mechanical system, where  an array of NV centers are  magnetically coupled to a  nanomechanical resonator. With the assistance of external microwave fields,
we can acquire collective interactions for NV spins with the form of the Lipkin-Meshkov-Glick (LMG) type \cite{LIPKIN1965188}
The LMG Hamiltonian can be  adiabatically steered from the isotropic type to the one-axis twisting one
by tuning the Rabi frequencies slowly enough to maintain the NV spins in the ground state.
The collective NV spins  undergo the ground-state transitions
that allows us to obtain the adiabatic channels between the initial separate ground state
and the final entangled ground state.
We investigate this adiabatic scheme with  analytical results
and numerical simulations for three different types of adiabatic transfer processes.
The results indicate  that we can acquire the Greenberger-Horne-Zeilinger (GHZ)-type
and the W-type entangled states for  NV spins with very high fidelity.
Compared to previous works,  this scheme is robust against practical noise and experimental imperfection because of adiabaticity.

\section{The setup}

\begin{figure}
\includegraphics[width=7cm]{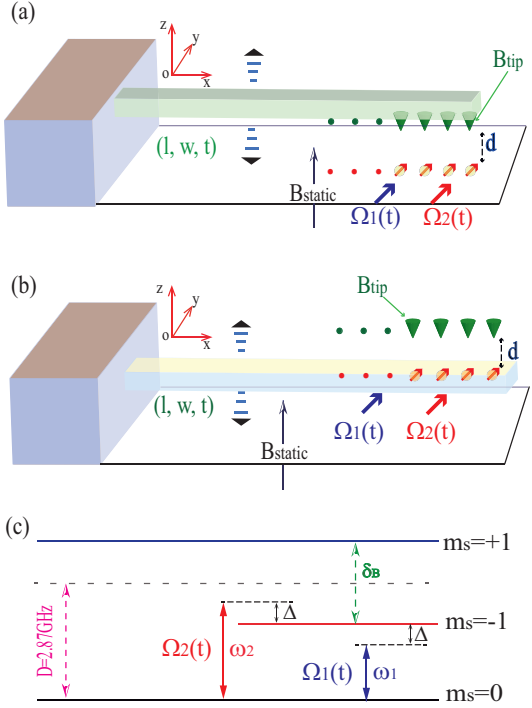}
\caption{\label{fig:wide}(Color online) The scheme diagrams.
(a) An array of equidistant magnetic tips are attached near the end of a cantilever  NAMR,
under which are $N$  distant NV centers with the same distance $d$.
In addition, two microwave fields are applied to drive the NV centers  between
the state $|m_{s}=0\rangle$ and the state $|m_{s}=-1\rangle$.
(b) Another feasible equivalent setup.
An array of NV centers are embedded equidistantly near  the end of a diamond NAMR,
above which are  $N$  magnetic tips with same distance $d$.
Two microwave fields are also applied to drive the transitions between
the state $|m_{s}=0\rangle$ and the state $|m_{s}=-1\rangle$ for the NV centers.
(c) Level diagram of the NV center ground triplet state
and the feasible transition channels.
The blue and red solid arrows indicate the two different microwave driving fields (with frequencies $\omega_{1}$, and $\omega_{2}$, and Rabi frequencies $\Omega_{1}$ and $\Omega_{2}$) applied between the state $|m_{s}=0\rangle$ and the state $|m_{s}=-1\rangle$.  }
\end{figure}

We consider the spin-mechanical setups as illustrated in Fig.~1(a) and (b).
The ground-state energy level structure of a single NV center is shown in Fig.~1(c).
The electronic ground triplet state $|m_{s}=0,\pm 1\rangle$ is the eigenstates of spin operator $\hat{S}_{z}$ with $\hat{S}_{z}|m_{s}\rangle=m_{s}|m_{s}\rangle$,
and the zero-field splitting between the degenerate
sublevels $|m_{s}=\pm 1\rangle$ and $|m_{s}=0\rangle $ is $D=2\pi \times 2.87$ GHz
\cite{PhysRevB.85.205203, Maze2011Properties, Doherty2013The}.
A homogeneous static magnetic field $B_{\text{static}}$ is used to remove the degenerate
states $|m_{s}=\pm 1\rangle$ with the Zeeman splitting $\hbar\delta _{B}=2g_{e}\mu
_{B}B_{\text{static}}$. In Fig.~1(a), the end of a cantilever  NAMR
with dimensions $(l,w,t)$ is attached with a row of equidistant magnet tips (size of $\sim100$ nm).
An array of NV centers are placed homogeneously and sparsely in the vicinity of the upper surface of the diamond sample,
which are placed  just under the magnet tips one-by-one with the same distance $d\sim25$ nm.
The motion of the cantilever attached with the magnet chip produces the time-dependent gradient magnetic field $\vec{B}_{j}(t)$,
with the fundamental frequency $\nu$ at the $j\text{-th}$ NV spin\cite{NatPhysWrachtrup, Albrecht2013Coupling,PhysRevB.79.041302,PhysRevA.80.022335,PhysRevA.96.023827}.
Meanwhile, we apply the dichromatic microwave driving fields polarized in the $x$ direction $B_{x}^{1,2}(t)$
with frequencies $\omega_{1}$ and $\omega_{2}$ to manipulate the NV centers' triple ground states.
To make sure that the NV centers are all strongly
and nearly equally coupled to the cantilever, we  restrict the tips within a small region near the end of the
cantilever. In this case, the number for
available NV centers is limited to at most 10 NV centers \cite{PhysRevB.79.041302,PhysRevA.80.022335}.
Moreover, we assume that the  dichromatic microwave fields drive the NV centers homogeneously,
because the microwave length is much larger than the size of the the cantilever.

Then for the single $j\text{-th}$ NV spin, we can obtain the Hamiltonian expressed as $(\hbar =1)$
\begin{eqnarray}\label{ME1}
H_{j}&=&D\hat{S}_{z}^{j2}+\frac{\delta_{B}}{2}\hat{S}_{z}^{j}\notag\\
&+&g_{e}\mu _{B}[\vec{B}_{j}(t)\cdot\hat{\vec{S}}^{j}+(B_{x}^{1}(t)+B_{x}^{2}(t))\hat{S}_{x}^{j}],
\end{eqnarray}
where $g_{e}\simeq 2$ is the land\'{e}
factor of NV center, $\mu _{B}=14 \text{GHz}/T$ is the Bohr magneton,
and $\hat{\vec{S}}^{j}\equiv(\hat{S}_{x}^{j},\hat{S}_{y}^{j},\hat{S}_{z}^{j})$ is the spin operator of the NV center.
As $\nu\ll D\pm\delta_{B}/2$, we can ignore the far-off resonant interactions between the spin  and the gradient magnetic fields along $x$ and $y$ directions. Then  we can obtain the Hamiltonian
\begin{eqnarray}\label{ME2}
H_{j}&\simeq&D\hat{S}_{z}^{j2}+\frac{\delta_{B}}{2}\hat{S}_{z}^{j}\notag\\
&+&g_{e}\mu _{B}[B_{z}^{j}(t)\hat{S}_{z}^{j}+(B_{x}^{1}(t)+B_{x}^{2}(t))\hat{S}_{x}^{j}].
\end{eqnarray}
We assume $B_{z}^{j}(t)\sim G_{j}\hat{z}\cos\nu t=G_{j}a_{0}(\hat{a}+\hat{a}^{\dag})\cos\nu t$, with $G_{j}$ the first order gradient magnetic field,
$\hat{a}$ and $\hat{a}^{\dag}$ the corresponding annihilation and creation operators,
and $a_{0}=\sqrt{\hbar/2m\nu}$ the zero field fluctuation for this resonator of mass $m$.
In the rotating frame at the frequency $\nu$,
\begin{eqnarray}\label{ME3}
H_{j}^{'}&\simeq&D\hat{S}_{z}^{j2}+\frac{\delta_{B}}{2}\hat{S}_{z}^{j}+\nu\hat{a}^{\dag}\hat{a}\notag\\
&+&\frac{1}{2}\lambda_{j}(\hat{a}+\hat{a}^{\dag})\hat{S}_{z}+g_{e}\mu _{B}[B_{x}^{1}(t)+B_{x}^{2}(t)]\hat{S}_{x}^{j},
\end{eqnarray}
where $\lambda_{j}=g_{e}\mu _{B}G_{j}a_{0}$ is the coupling constant between the $j\text{-th}$ NV center and the NAMR.
Taking $B_{x}^{1}(t)=B_{0}^{1}\cos\omega_{1}t$ and $B_{x}^{2}(t)=B_{0}^{2}\cos\omega_{2}t$,
we assume the frequencies of the two driving fields $\omega_{1}$ and $\omega_{2}$
are far off resonance with respect to the transition  between the states $|0\rangle$ and $|+1\rangle$.
Therefore, this allows us to isolate a two-level subsystem comprised by $\{|0\rangle,|-1\rangle\}$ for the single NV center.
For the $j\text{-th}$ NV spin, we can define $\hat{\sigma}_{z}^{j}\equiv(|-1\rangle _{j}\langle -1|-|0\rangle_{j}\langle 0|)$,
$\hat{\sigma}_{+}^{j}\equiv|-1\rangle _{j}\langle 0|$, and $\hat{\sigma}_{-}^{j}\equiv|0\rangle _{j}\langle -1|$.
Then we can obtain the Hamiltonian under the rotating-wave approximation,
\begin{eqnarray}\label{ME4}
H_{j}^{'}&=&\frac{\omega_{-}}{2}\hat{\sigma}_{z}^{j}+\nu\hat{a}^{\dag}\hat{a}+\frac{1}{2}\lambda_{j}(\hat{a}+\hat{a}^{\dag})\hat{\sigma}_{z}^{j}\notag\\
&+&\hat{\sigma}_{+}^{j}(\Omega_{1}e^{-i\omega_{1}t}+\Omega_{2}e^{-i\omega_{2}t})+H.c.,
\end{eqnarray}
where $\omega_{-}=D-\delta_{B}/2$ is the energy transition frequency between the level $|0\rangle$ and $|-1\rangle$,
and $\Omega_{1,2}=g_{e}\mu _{B}B_{0}^{1,2}/2$ are the dichromatic Rabi frequencies.

We can ignore the interactions between the adjacent NV centers, as long as the distance between the two adjacent NV spins is far enough.
Then we have  the total Hamiltonian for this hybrid system
\begin{eqnarray}\label{ME5}
H_{s}&=&\nu\hat{a}^{\dag}\hat{a}+\sum_{j=1}^{N}[\frac{\omega_{-}}{2}\hat{\sigma}_{z}^{j}+\frac{1}{2}\lambda_{j}(\hat{a}+\hat{a}^{\dag})\hat{\sigma}_{z}^{j}\notag\\
&+&\hat{\sigma}_{+}^{j}(\Omega_{1}e^{-i\omega_{1}t}+\Omega_{2}e^{-i\omega_{2}t})+H.c.].
\end{eqnarray}
The first and second items are the free Hamiltonian for the NAMR and NV centers,
the third item is the Hamiltonian for describing the interactions between the NV centers and the mechanical resonator,
and the last item describes the microwave  driving for the transition between $|0\rangle$ and $|-1\rangle$ of the NV spins.

We can also implement such a spin-mechanical setup by use of a suspended carbon nanotube resonator that carries dc current.
Recently, it  has been shown that the suspended carbon nanotube carrying dc current
can enable the strong coupling between  mechanical motion  and NV spins \cite{PhysRevLett.117.015502}.  This setup is
particularly suitable for the investigation of an array of NV centers coupled to a mechanical resonator.

Another equivalent  setup is illustrated in Fig.~1(b).
In this  system, a row of equidistant NV centers are set homogeneously and sparsely
in the vicinity of the upper surface near the end of the cantilever diamond NAMR.
An array of magnetic tips are fixed   above these NV centers one-by-one with the same distance $d$.
With the assistance of the static magnetic fields and  microwave driving fields,
we can also achieve the equivalent Hamiltonian for describing the interactions as the first setup shown in Fig.~1(a).

Owing to the variations in the size and spacing of the nanomagnets and NV centers,
the coupling $\lambda_{j}$ can not be the same for all of the NV centers.
There will be slight differences for each NV center, and this will
give rise to a degree of disorder in the system.
Here we define $\lambda_{j}=\lambda+\delta\lambda_{j}$ and $\eta_{j}=\eta+\delta\eta_{j}=(\lambda+\delta\lambda_{j})/\nu$,
where $|\delta\eta_{j}|=|\delta\lambda_{j}|/\nu \ll1$ is the disorder factor in this hybrid system \cite{PhysRevB.97.155133,PhysRevA.86.023837,PhysRevA.92.062305,PhysRevA.98.023628}.
Therefore, the Hamiltonian in Eq.~(\ref{ME5}) can be expressed as
\begin{eqnarray}\label{ME6}
H_{s}&=&\nu\hat{a}^{\dag}\hat{a}+\sum_{j=1}^{N}[\frac{\omega_{-}}{2}\hat{\sigma}_{z}^{j}+\frac{1}{2}(\lambda+\delta\lambda_{j})(\hat{a}+\hat{a}^{\dag})\hat{\sigma}_{z}^{j}\notag\\
&+&\hat{\sigma}_{+}^{j}(\Omega_{1}e^{-i\omega_{1}t}+\Omega_{2}e^{-i\omega_{2}t})+H.c.].
\end{eqnarray}
For conveniently, we define the collective spin operators for all of the NV centers as
$\hat{J}_{z}=\sum_{j=1}^{N}\hat{\sigma}_{z}^{j}/2$,
$\hat{J}_{+}=\sum_{j=1}^{N}\hat{\sigma}_{+}^{j}$, $\hat{J}_{-}=\sum_{j=1}^{N}\hat{\sigma}_{-}^{j}$,
and they also satisfy the angular momentum commutation relations
$[\hat{J}_{i},\hat{J}_{j}]=i\varepsilon_{ijk}\hat{J}_{k},[\hat{J}_{+},\hat{J}_{-}]=2\hat{J}_{z},[\hat{J}_{z},\hat{J}_{\pm }]=\pm\hat{J}_{\pm }$.
Therefore, Eq.~(\ref{ME6}) can be simplified as
\begin{eqnarray}\label{ME7}
H_{s}&=&\nu\hat{a}^{\dag}\hat{a}+\omega_{-}\hat{J}_{z}+\lambda(\hat{a}+\hat{a}^{\dag})\hat{J}_{z}\notag\\
&+&\hat{J}_{+}(\Omega_{1}e^{-i\omega_{1}t}+\Omega_{2}e^{-i\omega_{2}t})+H.c.\notag\\
&+&\sum_{j=1}^{N}\delta\lambda_{j}(\hat{a}+\hat{a}^{\dag})\frac{\hat{\sigma}_{z}^{j}}{2}.
\end{eqnarray}

First of all, we apply the unitary Schrieffer-Wolff transformation
$\hat{U}=e^{-i\hat{P}}$ to $H_{s}$,
where $\hat{P}\equiv i\eta(\hat{a}^{\dag}-\hat{a})\hat{J}_{z}$,
and $\eta=\lambda/\nu$ can be viewed as an effective Lamb-Dicke parameter for this solid-state system \cite{Albrecht2013Coupling, PhysRevX.6.041060, P2010A}.
Then we have $H_{s}^{'}\rightarrow \hat{U}H_{s}\hat{U}^{\dag}$.
\begin{eqnarray}\label{ME8}
H_{s}&=&\nu\hat{a}^{\dag}\hat{a}+\omega_{-}\hat{J}_{z}+\hat{J}_{+}e^{\eta(\hat{a}^{\dag}-\hat{a})}(\Omega_{1}e^{-i\omega_{1}t}+\Omega_{2}e^{-i\omega_{2}t})+H.c.\notag\\
&+&\sum_{j=1}^{N}\frac{1}{2}\delta\lambda_{j}(\hat{a}+\hat{a}^{\dag})\hat{\sigma}_{z}^{j}-\sum_{j=1}^{N}(M_{j}\hat{\sigma}_{z}^{j})\hat{J}_{z},
\end{eqnarray}
where $M_{j}=\eta\delta\lambda_{j}/4$ is the coefficient of Ising interactions.
$\hat{D}_{1}\equiv\sum_{j=1}^{N}\delta\lambda_{j}(\hat{a}+\hat{a}^{\dag})\hat{\sigma}_{z}^{j}/2$ and $\hat{D}_{2}\equiv\sum_{j=1}^{N}(M_{j}\hat{\sigma}_{z}^{j})\hat{J}_{z}$
are the  experimental disorder items  in our system. The first item $\hat{D}_{1}$ corresponds to the high frequency oscillating item,
and its effective influence on the system can be discarded because  $|\delta\lambda_{j}|\ll\nu$.
The second item $\hat{D}_{2}$ corresponds to the major disorder, whose effect  will be
discussed in Sec. \uppercase\expandafter{\romannumeral5}.
For simplicity, we first assume that $\lambda_{j}\simeq\lambda$, and then we can discard the item $\hat{D}_{2}$  because of $M_{j}\simeq 0$.
As a result, we can acquire the Hamiltonian without the disorder
\begin{eqnarray}\label{ME9}
H_{s}&\simeq&\nu\hat{a}^{\dag}\hat{a}+\omega_{-}\hat{J}_{z}\notag\\
&+&\hat{J}_{+}e^{\eta(\hat{a}^{\dag}-\hat{a})}(\Omega_{1}e^{-i\omega_{1}t}+\Omega_{2}e^{-i\omega_{2}t})+H.c.
\end{eqnarray}

Secondly, we assume that the resonator is cooled sufficiently with extremely low ambient temperature,
so that this hybrid system satisfies the Lamb-Dicke limit $(\overline{n}+1)\eta_{j}^{2}\ll 1$,
where $\overline{n}=1/(e^{\hbar\nu/k_{B}T}-1)$ is the average number of the phonon for this oscillation mode with  temperature $T$ \cite{PhysRevLett.74.4091,PhysRevLett.82.1835,PhysRevLett.87.230404, PhysRevLett.87.137902, PhysRevLett.90.133601}.
Applying the approximate relation $e^{\pm \eta(\hat{a}^{\dag}-\hat{a})}\simeq 1\pm \eta(\hat{a}^{\dag}-\hat{a})$ to Eq.~(\ref{ME9})
we can acquire the Hamiltonian in the interaction picture
\begin{eqnarray}\label{ME10}
\begin{aligned}
H_{\text{IP}}&\simeq\hat{J}_{+}e^{i\omega_{-} t}[1+\eta(\hat{a}^{\dag}e^{i\nu t}-\hat{a}e^{-i\nu t})]\\
&\times(\Omega_{1}e^{-i\omega_{1}t}+\Omega_{2}e^{-i\omega_{2}t})+H.c.
\end{aligned}
\end{eqnarray}%
We define the detuning as $\Delta \equiv\omega_{2}-\omega_{-}\simeq \omega_{-}-\omega_{1}$,
and assume the relations in this hybrid system $\nu\gg\lambda$, $|\Delta|\gg |\Omega_{1,2}|$, and $\{\nu,|\Delta|,|\Delta\pm\nu|\}\gg \eta|\Omega_{1,2}|$.
Since the NAMR stays in relative lower energy state,
we can eliminate the resonator mode  and
ignore the items for the energy shift caused by this mechanical oscillation mode.
Then we can get the effective Hamiltonian  as follow \cite{James2012Effective}
\begin{equation}\label{ME11}
H_{\text{eff}}\simeq\alpha(\varepsilon\beta_{1}\beta_{2}\hat{J}_{z}+\beta_{1}^{2}\hat{J}_{x}^{2}+\beta_{2}^{2}\hat{J}_{y}^{2}),
\end{equation}%
where the effective coefficients are
\begin{eqnarray}\label{ME12}
\alpha&=& 2\eta^{2}\Delta/(\Delta^{2}-\nu^{2}), \notag\\
\varepsilon&\simeq& 2/\alpha\Delta,\notag\\
\beta_{1}&=&\Omega_{1}-\Omega_{2},\notag\\
\beta_{2}&=&\Omega_{1}+\Omega_{2}.
\end{eqnarray}
According to Eq. (\ref{ME11}), the effective Hamiltonian for this system  evidently
corresponds to the general LMG model for describing the  collective interactions of $N$ spin-$1/2$ particles.

Therefore, in this solid-state system, we introduce the single NV center's decoherence factor
as the dephasing rate $\gamma_{\text{dep}}^{j}\sim 1/T_{2}$ to the master equation with the expression
\begin{equation}\label{ME13}
\dot{\hat{\rho}}=-i[H_{\text{eff}},\hat{\rho}]+\sum_{j=1}^{N}\gamma_{\text{dep}}^{j}D[\hat{\sigma}_{z}^{j}]\hat{\rho}.
\end{equation}%

\section{Generating entangled  states via adiabatic transitions}

\begin{table*}
\caption{\label{tab:table1} Different types of the LMG model with the physical parameters.}
\begin{ruledtabular}
\begin{tabular}{lllll}
The different LMG model   &  Physical parameters conditions   \\\hline
$H_{\text{isotropy}}=\alpha\beta^{2}(\varepsilon\hat{J}_{z}+\hat{\textbf{J}}^2-\hat{J}_{z}^2)$  & $\Omega_{1}\neq 0$, $\Omega_{2}= 0$, $\beta_{1}=\beta_{2}=\beta$, $|\Delta|\neq\nu$. \\
$H_{y}=\alpha\beta_{2}^{2}\hat{J}_{y}^{2}$  & $\Omega_{1}=\Omega_{2}$, $\beta_{1}=0$, $\beta_{2}\neq0$, $|\Delta|\neq\nu$. \\
$H_{x}=\alpha\beta_{1}^{2}\hat{J}_{x}^{2}$  & $\Omega_{1}=-\Omega_{2}$, $\beta_{1}\neq0$, $\beta_{2}=0$, $|\Delta|\neq\nu$.\\
\end{tabular}
\end{ruledtabular}
\end{table*}

\begin{table*}
\caption{\label{tab:table2}The ground states for different types of the LMG model.}
\begin{ruledtabular}
\begin{tabular}{lllcc}
Hamiltonian for different types of LMG model &   The ground state \\\hline
 $H_{\text{isotropy}}=\alpha\beta^{2}(\varepsilon\hat{J}_{z}+\hat{\textbf{J}}^2-\hat{J}_{z}^2)$ ($|\varepsilon|> N$ and $\alpha<0$) & $|\uparrow\uparrow\cdots\uparrow\rangle (\varepsilon>0)$
 or $|\downarrow\downarrow\cdots\downarrow\rangle (\varepsilon<0)$\\
 $H_{\text{isotropy}}=\alpha\beta^{2}(\varepsilon\hat{J}_{z}+\hat{\textbf{J}}^2-\hat{J}_{z}^2)$ ($|\varepsilon|> N$ and $\alpha>0$) & $|\uparrow\uparrow\cdots\uparrow\rangle (\varepsilon<0)$
 or $|\downarrow\downarrow\cdots\downarrow\rangle (\varepsilon>0)$\\
 $H_{y}=\alpha\beta_{2}^{2}\hat{J}_{y}^{2}$ ($\alpha<0$) & $|++\cdots+\rangle_{y}$ or $|--\cdots-\rangle_{y}$\\
 $H_{y}=\alpha\beta_{2}^{2}\hat{J}_{y}^{2}$ ($\alpha>0$) & $|(+)^{N/2}(-)^{N/2}\rangle_{y}$ (N is even) and $|(+)^{(N\pm1)/2}(-)^{(N\mp1)/2}\rangle_{y}$ (N is odd)\\
 $H_{x}=\alpha\beta_{1}^{2}\hat{J}_{x}^{2}$ ($\alpha<0$) & $|++\cdots+\rangle_{x}$ or $|--\cdots-\rangle_{x}$\\
 $H_{x}=\alpha\beta_{1}^{2}\hat{J}_{x}^{2}$ ($\alpha>0$) & $|(+)^{N/2}(-)^{N/2}\rangle_{x}$ (N is even) and $|(+)^{(N\pm1)/2}(-)^{(N\mp1)/2}\rangle_{x}$ (N is odd)\\
\end{tabular}
\end{ruledtabular}
\end{table*}

The parameters $\alpha$, $\varepsilon$, $\beta_{1}$ and $\beta_{2}$ in Eq. (\ref{ME12}) can be controlled by
adjusting the relevant parameters such as  the detunings $\Delta$, Rabi frequencies $\Omega_{1,2}$,
and coupling coefficients $\lambda$. We can get several special forms of the LMG model
by tuning these parameters and make a concise list in TABLE. \uppercase\expandafter{\romannumeral1}.

The LMG model was first proposed by H. J. Lipkin, M. Meshkov and A. J. Glick
for describing the monopole-monopole interactions in  nuclear physics \cite{LIPKIN1965188}.
In order to explore new physics from this LMG type interaction,
so far, a great deal of theoretical schemes are proposed
for simulating this kind of interaction with different systems,
such as the ion-trap scheme \cite{PhysRevLett.82.1835}, the cavity QED scheme \cite{PhysRevLett.100.040403},
and the hybrid solid-state qubit scheme \cite{PhysRevA.96.062333,Nori113020}.
The LMG type Hamiltonian possesses the particular symmetry under the exchange of particles.
Especially, the isotopic ferromagnetic LMG model and the simple (one-axis twisting)
ferromagnetic or antiferromagnetic LMG model can be solved exactly. Here
we study the ground states for these different types of the LMG model and make a brief list in TABLE. \uppercase\expandafter{\romannumeral2}.

\begin{figure}
\includegraphics[width=8.5cm]{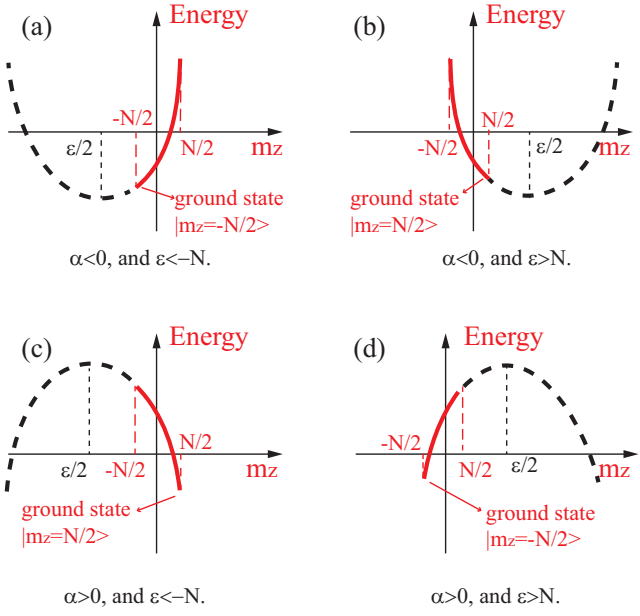}
\caption{\label{fig:wide}(Color online) The analysis graphics for the ground state of the
isotropy LMG Hamiltonian $H_{\text{isotropy}}$ with four different conditions:
(a) and (b) are for the negative coefficient $\alpha<0$ with $\varepsilon < -N$ and $\varepsilon > N$, while
(c) and (d) are for the positive coefficient $\alpha>0$ with $\varepsilon < -N$ and $\varepsilon > N$.}
\end{figure}

Let's make a brief discussion on these different types of the LMG model.
When we choose the experimental parameters as $\Omega_{1}\neq 0$, $\Omega_{2}= 0$,
$\beta_{1}=\beta_{2}=\beta$ and $|\Delta|\neq\nu$
according to the first row in TABLE. \uppercase\expandafter{\romannumeral1},
we can get the isotropic Hamiltonian with the expression
\begin{eqnarray}\label{ME14}
H_{\text{isotropy}}&=\alpha\beta^{2}(\varepsilon\hat{J}_{z}+\hat{J}_{x}^{2}+\hat{J}_{y}^{2})\notag\\
&=\alpha\beta^{2}(\varepsilon\hat{J}_{z}+\hat{\textbf{J}}^2-\hat{J}_{z}^2),
\end{eqnarray}%
where the coefficients are $\alpha= 2\eta^{2}\Delta/(\Delta^{2}-\nu^{2})$, $\varepsilon=2/\alpha\Delta$ and $\beta=\Omega_{1}$.
In which, the collective spins operator is
$\hat{\textbf{J}}^2=\hat{J}_{x}^{2}+\hat{J}_{y}^{2}+\hat{J}_{z}^{2}=J(J+1)$,
with the maximum total angular momentum $J= N/2$ ($J$ is integer or half integer number).
This  isotropic LMG Hamiltonian can be solved exactly in the representation of $\hat{J}_{z}$ because of the relations
$H_{\text{isotropy}}|m_{z}\rangle=\alpha\beta^{2}[\varepsilon m_{z}-m_{z}^2+J(J+1)]|m_{z}\rangle$,
where $m_{z}\in \{-J,-J+1,\cdots, J-1,J\}$, and $\{|m_{z}\rangle\}$ is the eigenstates of $\hat{J}_{z}$.
In order to describe the physics more visually,  in Fig. 2 we show the analysis graphics for the ground states of the
isotropy LMG Hamiltonian $H_{\text{isotropy}}$ in different conditions.

As shown in Fig.~2(a) and (b), when $\alpha< 0$, $|\varepsilon|> N$, and because of the symmetry breaking $\varepsilon\neq0$,
there must be a unique ground state for Eq. (\ref{ME14}), and the result is shown in the first row of TABLE.\uppercase\expandafter{\romannumeral2},
with $|m_{z}=N/2\rangle\equiv|\uparrow\uparrow\cdots\uparrow\rangle$ $(\varepsilon>0)$ and  $|m_{z}=-N/2\rangle\equiv|\downarrow\downarrow\cdots\downarrow\rangle$ $(\varepsilon<0)$.
Here $\vert \uparrow\rangle\equiv \vert -1\rangle$ and $\vert \downarrow\rangle\equiv \vert 0\rangle$.
On the other hand, if we set $\alpha> 0$ and $|\varepsilon|> N$,
we can also give a convincing interpretation of the ground state for $H_{\text{isotropy}}$ according to Fig.~2(c) and (d).
Then we can get the unique ground state for Eq. (\ref{ME14}) in the second row of TABLE.\uppercase\expandafter{\romannumeral2}, with $|m_{z}=N/2\rangle\equiv|\uparrow\uparrow\cdots\uparrow\rangle$ $(\varepsilon<0)$ and $|m_{z}=-N/2\rangle\equiv|\downarrow\downarrow\cdots\downarrow\rangle$ $(\varepsilon>0)$.

By setting the parameters as the second and third rows in TABLE.\uppercase\expandafter{\romannumeral1},
we can also get the simple (one-axis twisting) LMG Hamiltonian
\begin{equation}\label{ME15}
H_{y}=\alpha\beta_{2}^{2}\hat{J}_{y}^{2},
\end{equation}%
where the parameters are $\Omega_{1}=\Omega_{2}$, $\beta_{1}=0$, $\beta_{2}\neq0$, $|\Delta|\neq\nu$,
and $\alpha= 2\eta^{2}\Delta/(\Delta^{2}-\nu^{2})$.
\begin{equation}\label{ME16}
H_{x}=\alpha\beta_{1}^{2}\hat{J}_{x}^{2},
\end{equation}%
with the parameters $\Omega_{1}=-\Omega_{2}$, $\beta_{1}\neq0$, $\beta_{2}=0$, $|\Delta|\neq\nu$,
and $\alpha= 2\eta^{2}\Delta/(\Delta^{2}-\nu^{2})$.

According to Eq.~(\ref{ME15}) and Eq.~(\ref{ME16}),
when we change the sign of $\alpha$ from the negative value to the positive one,
the collective spin system  correspondingly undergoes the phase transition
from the ferromagnetic interactions (FI) to the antiferromagnetic interactions (AFI).
These transitions can also lead to the collective NV spins' ground-state transitions
 shown in TABLE. \uppercase\expandafter{\romannumeral2}.
When $\alpha<0$, Eqs. (\ref{ME15}) and (\ref{ME16}) are the Hamiltonians for describing the FI,
whose ground states are double degenerate ones according to the third and the fifth rows in TABLE. \uppercase\expandafter{\romannumeral2},
with the expressions $|m_{x,y}=N/2\rangle\equiv|++\cdots+\rangle_{x,y}$ and $|m_{x,y}=-N/2\rangle\equiv|--\cdots-\rangle_{x,y}$.
Here $\vert \pm\rangle_{x}\equiv (\vert \uparrow\rangle\pm\vert \downarrow\rangle)/\sqrt{2}$
and $\vert \pm\rangle_{y}\equiv (\vert \uparrow\rangle\pm i\vert \downarrow\rangle)/\sqrt{2}$.
On the contrary, if we set $\alpha>0$, we can have the Hamiltonian for AFI,
and obtain the ground states  corresponding to the fourth and the sixth rows in TABLE. \uppercase\expandafter{\romannumeral2}, i.e.,
$|m_{x,y}=0\rangle\equiv|(+)^{N/2}(-)^{N/2}\rangle_{x,y}$ ($N$ is even) and $|m_{x,y}=\pm \frac{1}{2}\rangle\equiv|(+)^{(N\pm1)/2}(-)^{(N\mp1)/2}\rangle_{x,y}$ ($N$ is odd).

In this work, we focus on the generation of the multiparticle entangled states through
adiabatically steering the Hamiltonian from the isotropic type $H_{\text{isotopic}}$ to the one-axis twisting one $H_{y}$.
The essential criteria for this scheme is that we need to keep all  spins in the ground states during the dynamical evolution process.
It is necessary  to determine the slowly varying functions of the Rabi frequencies $\Omega_{1,2}(t)$ versus the evolution time.
Then we tune the parameters $\Omega_{1,2}(t)$ slowly enough to maintain the adiabatic conditions $\tau\gg\hbar/\Delta E$,
in which $\tau$ is the characteristic time for the transfer process,
and $\Delta E$ is the energy difference between the ground state and the next excited state.
According to the discussion in Ref. \cite{PhysRevLett.82.1835}, this adiabaticity constraints will not change as we increase the number of
particles up to 50. In our scheme, the number of the NV centers have been limited in
N ¡Ü 10, as a result, the adiabaticity constraints will be valid.
We consider  three different  schemes: case \uppercase\expandafter{\romannumeral1}, $\alpha<0$, $\varepsilon>0$; case \uppercase\expandafter{\romannumeral2}, $\alpha>0$, $\varepsilon>0$  and $N$ is odd; case \uppercase\expandafter{\romannumeral3},
$\alpha>0$, $\varepsilon>0$  and $N$ is even.
These three different types of adiabatic processes are shown in Fig.~3.
\begin{figure}
\includegraphics[width=8.5cm]{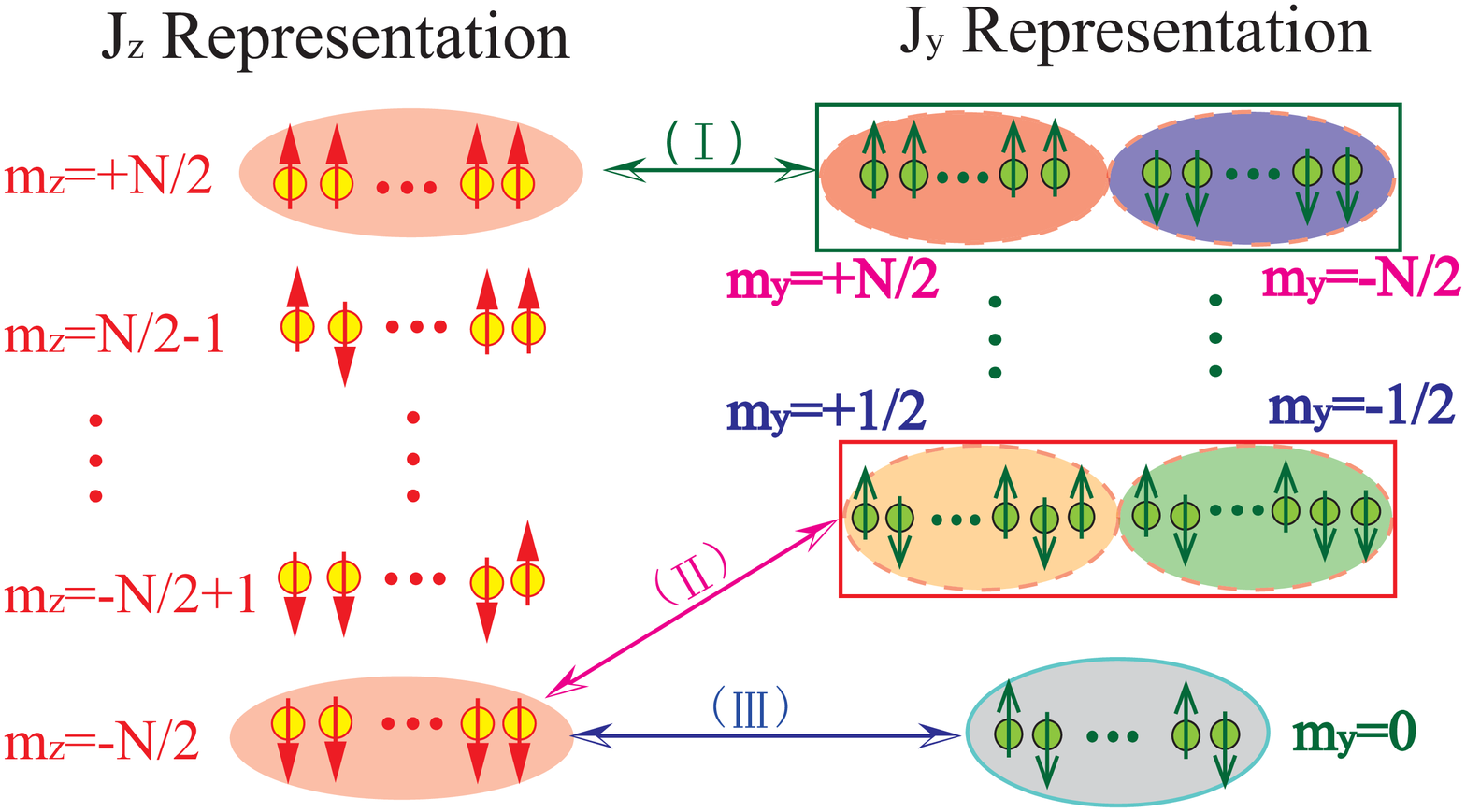}
\caption{\label{fig:wide}(Color online) The diagram for three different adiabatic transition schemes (case \uppercase\expandafter{\romannumeral1}, case \uppercase\expandafter{\romannumeral2},  and case \uppercase\expandafter{\romannumeral3}) between the initial ground states (left)
and the final ground states (right) for all of the NV spins. In case \uppercase\expandafter{\romannumeral1}, the transition is $|m_{z}=N/2\rangle\xrightarrow{\text{(\uppercase\expandafter{\romannumeral1})}} |\phi_{\text{(\uppercase\expandafter{\romannumeral1})}}\rangle$,
with the parameters $\alpha<0$ and $\varepsilon>N$; in case \uppercase\expandafter{\romannumeral2}, the transition is $|m_{z}=-N/2\rangle\xrightarrow{\text{(\uppercase\expandafter{\romannumeral2})}} |\phi_{\text{(\uppercase\expandafter{\romannumeral2})}}\rangle$,
with the parameters $\alpha>0$, $\varepsilon>N$ and odd number of NV spins;  for case \uppercase\expandafter{\romannumeral3}, the transition is $|m_{z}=-N/2\rangle\xrightarrow{\text{(\uppercase\expandafter{\romannumeral3})}} |\phi_{\text{(\uppercase\expandafter{\romannumeral3})}}\rangle$,
with the parameters $\alpha>0$, $\varepsilon>N$ and even number of NV spins.}
\end{figure}

For case I, we set the coupling parameters to satisfy $\alpha<0$, $\varepsilon>N$, $\Omega_{1(\text{inital})}=\Omega_{1(\text{i})}$, $\Omega_{2(\text{inital})}=\Omega_{2(\text{i})}=0$ and $\beta_{1}=\beta_{2}=\beta$,
and assume that Eq. (\ref{ME14}) is the initial Hamiltonian in this hybrid quantum system,
which corresponds to the first row in TABLE.~\uppercase\expandafter{\romannumeral1}.
According to Fig.~2(b) and the first row in TABLE.~\uppercase\expandafter{\romannumeral2},
we can analytically achieve the unique initial ground states $|m_{z}=N/2\rangle$ for Eq. (\ref{ME14}),
which is the separable multiparticle state without any entanglement.
With the adiabatic transfer process $\Omega_{1(\text{i})}\xrightarrow{\tau}\Omega_{\text{f}}$
and $\Omega_{2(\text{i})}\xrightarrow{\tau}\Omega_{\text{f}}$,
we can transform the LMG Hamiltonian from Eq.~(\ref{ME14}) into Eq.~(\ref{ME15}).
As a result, we can achieve the adiabatic transfer process $\alpha\beta^{2}(\varepsilon\hat{J}_{z}+\hat{\textbf{J}}^2-\hat{J}_{z}^2)\xrightarrow{\text{Adiabatic}} \alpha\beta_{2}^{2}\hat{J}_{y}^{2}$ in this hybrid quantum system.
Moreover, since the Hamiltonian for this type of transition corresponds to the FI,
there is no need to discuss the odevity of the number of NV centers.
Owing to the particular symmetry of the exchange of particles for this kind of LMG-type interactions as Eqs. (\ref{ME14}) and (\ref{ME15}),
we can get the adiabatic ground-state transfer between the initial disentangled ground state and the final target entangled ground state,
\begin{equation}\label{ME17}
|m_{z}=\frac{N}{2}\rangle\xrightarrow{\text{(\uppercase\expandafter{\romannumeral1})}}\\
|\phi_{\text{(\uppercase\expandafter{\romannumeral1})}}\rangle\equiv\frac{1}{\sqrt{2}}[|m_{y}=\frac{N}{2}\rangle+e^{i\pi J}|m_{y}=-\frac{N}{2}\rangle].
\end{equation}%
Here $|\phi_{(\text{\uppercase\expandafter{\romannumeral1}})}\rangle$   corresponds to the $N$-particle GHZ-type entangled state in the $\hat{J}_{y}$ representation,
and the total number of spins $N$ can sensitively influence on the entanglement due to the phase factor $e^{i\pi J}$. For example, when  $N=4$,
we can have $|\phi_{\text{(\uppercase\expandafter{\romannumeral1})}}\rangle=\frac{1}{\sqrt{2}}(|++++\rangle_{y}+|----\rangle_{y})$.

For another case, according to the first row in TABLE. \uppercase\expandafter{\romannumeral1}
and the second row in TABLE. \uppercase\expandafter{\romannumeral2},
we set the parameters as $\alpha>0$, $\varepsilon>N$, $\Omega_{1(\text{inital})}=\Omega_{1(\text{i})}$,
$\Omega_{2(\text{inital})}=\Omega_{2(\text{i})}=0$ and $\beta_{1}=\beta_{2}=\beta$.
Therefore, for Eq.~(\ref{ME14}), the initial ground state can be  expressed as $|m_{z}=-N/2\rangle$, which is also plotted in Fig.~2(d).
Taking advantage of the same adiabatic transfer process $\Omega_{1(\text{i})}\xrightarrow{\tau}\Omega_{\text{f}}$ and $\Omega_{2(\text{i})}\xrightarrow{\tau}\Omega_{\text{f}}$,
we can also achieve the transfer process $\alpha\beta^{2}(\varepsilon\hat{J}_{z}+\hat{\textbf{J}}^2-\hat{J}_{z}^2)\xrightarrow{\text{Adiabatic}} \alpha\beta_{2}^{2}\hat{J}_{y}^{2}$ to prepare the entangled ground states.
Since $\alpha>0$, the initial Hamiltonian is the form of FI, but the final Hamiltonian is the type of AI.
In this adiabatic transfer process, the odevity of $N$ needs to be distinguished between odd (case \uppercase\expandafter{\romannumeral2}) and even (case \uppercase\expandafter{\romannumeral3}), as illustrated in Fig.~3.

In case \uppercase\expandafter{\romannumeral2}, $\alpha>0$ and the total number of NV spins $N$ is odd.
The ground states for the final Hamiltonian $H_{y}=\alpha\beta_{2}^{2}\hat{J}_{y}^{2}$ are two generate ground states as $|m_{y}=\pm1/2\rangle$,
which are both maximally entangled states. Moreover, owing to the symmetrical interactions for exchanging the particles,
we can get the second adiabatic transition process in case \uppercase\expandafter{\romannumeral2},
\begin{equation}\label{ME18}
|m_{z}=-\frac{N}{2}\rangle\xrightarrow{\text{(\uppercase\expandafter{\romannumeral2})}}
|\phi_{\text{(\uppercase\expandafter{\romannumeral2})}}\rangle\equiv\frac{1}{\sqrt{2}}[|m_{y}=\frac{1}{2}\rangle+i|m_{y}=-\frac{1}{2}\rangle],
\end{equation}%
where $|\phi_{(\text{\uppercase\expandafter{\romannumeral2}})}\rangle$ is the $N$-particle W-type maximally entangled state in the $\hat{J}_{y}$ representation,
with the total angular momentum $J=1/2$ for all the spins. Similarly, when $N=3$, we can obtain $|m_{y}=1/2\rangle=\frac{1}{\sqrt{3}}(|+-+\rangle_{y}+|-++\rangle_{y}+|++-\rangle_{y})$ and
$|m_{y}=-1/2\rangle=\frac{1}{\sqrt{3}}(|--+\rangle_{y}+|-+-\rangle_{y}+|+--\rangle_{y})$.
Evidently, $|m_{y}=\pm1/2\rangle$ are both the W-type maximally entangled state in the $\hat{J}_{y}$ representation.

For case \uppercase\expandafter{\romannumeral3}, $\alpha>0$ and the total number of NV spins $N$ is even.
With the final Hamiltonian $H_{y}=\alpha\beta_{2}^{2}\hat{J}_{y}^{2}$,
we have the unique nondegenerate ground state as $|m_{y}=0\rangle\equiv|(+)^{N/2}(-)^{N/2}\rangle_{y}$,
which is also the W-type maximally entangled state.
Then we obtain the third adiabatic transition process in case \uppercase\expandafter{\romannumeral3},
\begin{equation}\label{ME19}
|m_{z}=-\frac{N}{2}\rangle\xrightarrow{\text{(\uppercase\expandafter{\romannumeral3})}}
|\phi_{\text{(\uppercase\expandafter{\romannumeral3})}}\rangle\equiv|m_{y}=0\rangle,
\end{equation}%
where $|\phi_{(\text{\uppercase\expandafter{\romannumeral3}})}\rangle$ is also the
$N$-particle W-type maximally entangled state in the $\hat{J}_{y}$ representation,
with the total angular momentum $J=0$ for all the NV spins. When $N=4$, we have the target state as $|\phi_{(\text{\uppercase\expandafter{\romannumeral3}})}\rangle=\frac{1}{\sqrt{6}}(|++--\rangle_{y}+|+-+-\rangle_{y}
+|+--+\rangle_{y}+|-++-\rangle_{y}+|-+-+\rangle_{y}+|--++\rangle_{y})$, which   corresponds to the
four-particle W-type maximally entangled state.

\begin{figure}[t]
\includegraphics[width=11.5cm]{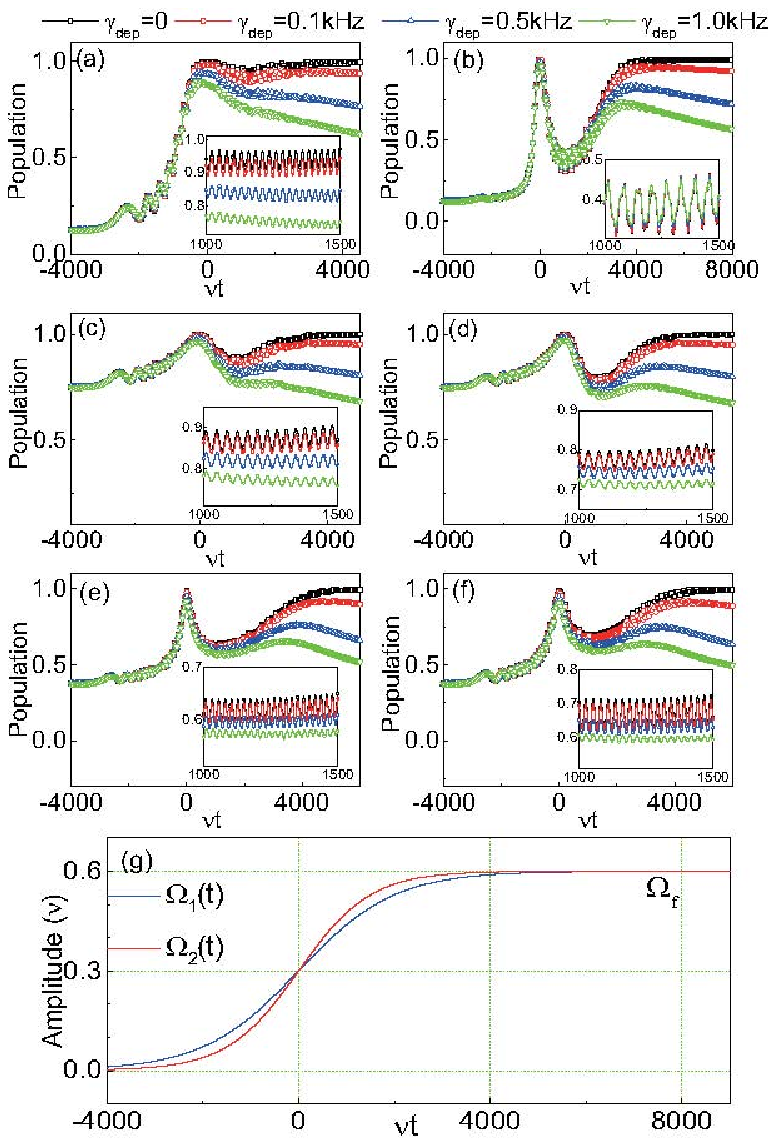}
\caption{\label{fig:wide}(Color online) The dynamical evolution for the population of the target entangled states
$|\phi_{(\text{\uppercase\expandafter{\romannumeral1}})}\rangle$, $|\phi_{(\text{\uppercase\expandafter{\romannumeral2}})}\rangle$, and $|\phi_{(\text{\uppercase\expandafter{\romannumeral3}})}\rangle$
in the adiabatic transfer scheme  for case \uppercase\expandafter{\romannumeral1}, case \uppercase\expandafter{\romannumeral2}, and case \uppercase\expandafter{\romannumeral3}, correspondingly,
with the slowly varying Rabi frequencies $\Omega_{1}(t)=0.3\nu[1+\tanh(\nu t/2000)]$ and $\Omega_{2}(t)=0.3\nu[1+\tanh(\nu t/1500)]$ as illustrated in (g).
For case \uppercase\expandafter{\romannumeral1}, (a) $\lambda=0.1\nu$ and $|\Delta|=1.1\nu$,  and  (b) $\lambda=0.05\nu$ and $|\Delta|=1.1\nu$;
for case \uppercase\expandafter{\romannumeral2}, (c) $\lambda=0.1\nu$ and $|\Delta|=1.1\nu$, and (d) $\lambda=0.1\nu$ and $|\Delta|=0.9\nu$;
for case \uppercase\expandafter{\romannumeral3},  (e) $|\Delta|=1.1\nu$ and $\lambda=0.1\nu$, and  (f) $|\Delta|=0.9\nu$ and $\lambda=0.1\nu$. }
\end{figure}

\section{Numerical simulations}
To confirm our theoretical schemes discussed above,
we assume that the frequency of the NAMR is about $\nu/2\pi=10$ MHz
and the dephasing rates for all the NV spins are homogeneous
$\gamma_{\text{dep}}^{j}\simeq\gamma_{\text{dep}}$.
Then we make the numerical simulations through solving the master equation (\ref{ME13}),
and display the results for different cases in Fig. 4.
In these numerical simulations, we have set the dephasing rate $\gamma_{\text{dep}}/2\pi$ respectively as $0$ kHz, $0.1$ kHz, $0.5$ kHz, and $1.0$ kHz.

The adiabatic ground-state transfer process for case \uppercase\expandafter{\romannumeral1} is plotted in Fig.~4(a) and (b).
We have assumed that the NV spins are initially prepared in the ground state $|m_{z}=N/2\rangle$ and set the number of NV spins as $N=4$.
This adiabatic process  corresponds to the transition of the FI LMG model
between the isotropy type and the one-axis twisting type as shown in Fig.~3.
We apply the slowly varying dichromatic microwave fields with the Rabi frequencies $\Omega_{1}(t)$ and $\Omega_{2}(t)$ according to Fig.~4(g),
and obtain the dynamical evolution of the population for the target state $|\phi_{(\text{\uppercase\expandafter{\romannumeral1}})}\rangle$.

In Fig.~4(a), the detuning is $|\Delta|=1.1\nu$ and the coupling is $\lambda\sim0.1\nu$,
while  in Fig.~4(b) the detuning is $|\Delta|=1.1\nu$ and the coupling is $\lambda\sim0.05\nu$.
We find that the collective NV spins will be transferred to the  GHZ state
$|\phi_{(\text{\uppercase\expandafter{\romannumeral1}})}\rangle$  at the time $t \sim 4000/\nu$.
When  the coupling strength between the NV centers and the NAMR decreases,
as shown in Fig.~4(b), the time for reaching the target state will be much longer.
Furthermore, we find that in ideal conditions the system can be steered into the target GHZ state $|\phi_{(\text{\uppercase\expandafter{\romannumeral1}})}\rangle$
with a fidelity equal to unity. However, when the spin dephasing effect is taken into account,
the population in the target state decreases.

For case \uppercase\expandafter{\romannumeral2} and case \uppercase\expandafter{\romannumeral3}, the adiabatic state transfer schemes
correspond to the transitions from the FI to the AFI as shown in Fig.~3.
In these cases, the target states depend on the odd or even number of the NV spins.
Therefore, for case \uppercase\expandafter{\romannumeral2},
we have set the odd number of NV spins as $N=3$ and assumed that the NV spins are initially prepared in the ground state $|m_{z}=-N/2\rangle$.
With the assistance of the slowly varying dichromatic microwave fields according to Fig.~4(g),
we can acquire the dynamical evolution of the population for the target state
$|\phi_{(\text{\uppercase\expandafter{\romannumeral2}})}\rangle$ in our numerical simulation.

As shown in Fig.~4(c)-(d), we set $\lambda=0.1\nu$, $|\Delta|=1.1\nu$ in Fig. 4(c), and $\lambda=0.1\nu$, $|\Delta|=0.9\nu$ in Fig. 4(d).
We find that the collective NV spins will be transferred to the W-type entangled ground state
$|\phi_{(\text{\uppercase\expandafter{\romannumeral2}})}\rangle$ when $t \sim 4000/\nu$  under different detunings.
We can also find that the population of the target W-state $|\phi_{(\text{\uppercase\expandafter{\romannumeral2}})}\rangle$
can reach unity in ideal conditions, but less than unity in real conditions because of the dephasing effect.

While for case \uppercase\expandafter{\romannumeral3},
we have set the even number of NV spins as $N=4$ and assumed that the NV spins are also initially prepared in the ground state $|m_{z}=-N/2\rangle$.
With the assistance of the identical dichromatic microwave fields according to Fig. 4(g),
we can also obtain the dynamical evolution of the population for the target state $|\phi_{(\text{\uppercase\expandafter{\romannumeral3}})}\rangle$,
as illustrated in Fig. 4(e) and (f). The parameters are chosen the same as those in Fig. 4(c) and (d).
Obviously, in spite of the different detunings the collective four NV spins will be transferred to the
W-type entangled ground state $|\phi_{(\text{\uppercase\expandafter{\romannumeral3}})}\rangle$ at the time  $t \sim 4000/\nu$.
We also find that the NV spins can be steered into the target W-type state $|\phi_{(\text{\uppercase\expandafter{\romannumeral3}})}\rangle$
with a fidelity equal to unity in ideal conditions. However, when the spin dephasing effect is taken into account, the population in the target state decreases.

\section{Experimental imperfections}

We now discuss the experimental imperfections. In this scheme, the experimental imperfections are mainly the physical disorder and the dispersion of the control parameters. Owing to the variations in the size and spacing of the magnetic tips and NV centers,
and according to the discussion in Sec.\uppercase\expandafter{\romannumeral2},
the physical disorder is mainly caused by the inhomogeneous coupling $\lambda_{j}$ between the NV centers and magnetic tips.
We can make  numerical simulations and display the effect of different disorder distributions on our scheme \cite{PhysRevB.97.155133,PhysRevA.86.023837,PhysRevA.92.062305,PhysRevA.98.023628}.
We set $\lambda=0.1\nu$, $|\Delta|=0.9\nu$, and the slowly varying Rabi frequencies
$\Omega_{1}(t)=0.3\nu[1+\tanh(\nu t/2000)]$ and $\Omega_{2}(t)=0.3\nu[1+\tanh(\nu t/1500)]$.
In Fig.~5, we plot the transfer efficiency under different disorder distributions.
Moreover, according to the different values of $|\delta\lambda_{j}|$,
we consider four different cases (disorder-(a,b,c,d)) in Fig.~5:
$|\delta\lambda_{j}|\leq\lambda\times5\%$ in Fig.~5(a),
$|\delta\lambda_{j}|\leq\lambda\times10\%$ in Fig.~5(b),
$|\delta\lambda_{j}|\leq\lambda\times20\%$ in Fig.~5(c),
and $|\delta\lambda_{j}|\leq\lambda\times30\%$ in Fig.~5(d).

In our simulations, we take four spins as an example. In Fig.~5,  we choose three different distributions for each disorder case.
According to the numerical simulations as shown in Fig.~5 (a)-(d),
we find that the collective NV spins will be transferred to the target
ground state when the disorder is about $\pm5\%\sim\pm30\%$ of $\lambda$,
and this transfer process is  unaffected by these kinds of disorder.

\begin{figure}[t]
\includegraphics[width=9cm]{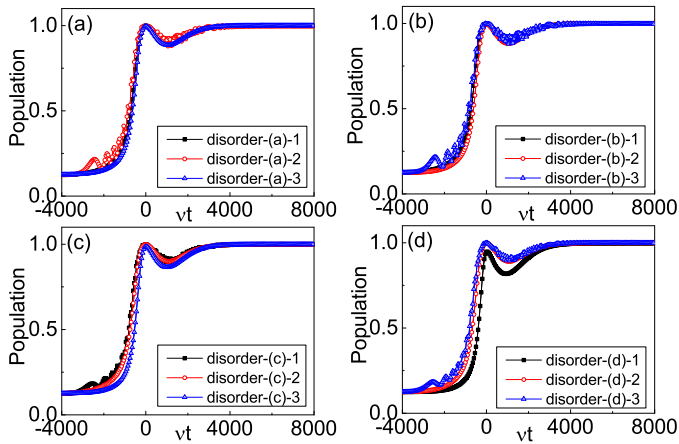}
\caption{\label{fig:wide}(Color online) The dynamical evolution for the population of the target entangled ground state
$|\phi_{(\text{\uppercase\expandafter{\romannumeral1}})}\rangle$ under different disorder distributions,
with the parameters $\lambda=0.1\nu$, and $|\Delta|=0.9\nu$. (a) $|\delta\lambda_{j}|\leq\lambda\times5\%$,
disorder-(a)-1: \{$-0.05\lambda$, $0.05\lambda$, $0.04\lambda$, $0.05\lambda$\},
disorder-(a)-2: \{$-0.05\lambda$, $0.04\lambda$, $-0.05\lambda$, $0.05\lambda$\},
and disorder-(a)-3: \{$0.05\lambda$, $0.04\lambda$, $0.05\lambda$, $0.04\lambda$\}.
(b) $|\delta\lambda_{j}|\leq\lambda\times10\%$,
disorder-(b)-1: \{$0.1\lambda$, $-0.05\lambda$, $0.08\lambda$, $-0.04\lambda$\},
disorder-(b)-2: \{$0.05\lambda$, $0.09\lambda$, $0.07\lambda$, $0.01\lambda$\},
and disorder-(b)-3: \{$-0.05\lambda$, $-0.03\lambda$, $-0.02\lambda$, $0.1\lambda$\}.
(c) $|\delta\lambda_{j}|\leq\lambda\times20\%$,
disorder-(c)-1: \{$-0.2\lambda$, $-0.01\lambda$, $0.15\lambda$, $0.07\lambda$\},
disorder-(c)-2: \{$-0.12\lambda$, $-0.15\lambda$, $0.2\lambda$, $-0.1\lambda$\},
and disorder-(c)-3: \{$0.2\lambda$, $0.05\lambda$, $0.11\lambda$, $-0.01\lambda$\}.
(d) $|\delta\lambda_{j}|\leq\lambda\times30\%$,
disorder-(d)-1: \{$0.3\lambda$, $0.2\lambda$, $0.1\lambda$, $-0.01\lambda$\},
disorder-(d)-2: \{$-0.1\lambda$, $-0.2\lambda$, $0.3\lambda$, $0.15\lambda$\},
and disorder-(d)-3: \{$-0.2\lambda$, $0.3\lambda$, $-0.1\lambda$, $0.01\lambda$\}.
}
\end{figure}

The dispersion caused by the experimental control parameters
is another experimental imperfection.
In this scheme, the dispersion mainly results from the dichromatic
slowly varying Rabi frequencies $\Omega_{1}(t)$ and $\Omega_{2}(t)$.
We assume $\Omega_{1}(t)=(\zeta+\delta\zeta_{1})[1+\tanh(\nu t/2000)]$
and $\Omega_{2}(t)=(\zeta+\delta\zeta_{2})[1+\tanh(\nu t/1500)]$,
with the average value $\zeta=0.3\nu$,
and the dispersions $|\delta\zeta_{1,2}|\leq\zeta\times10\%$.
By setting $\lambda=0.1\nu$, $|\Delta|=0.9\nu$, we plot the dynamical evolution
for the ground-state transfer  efficiency in Fig.~6 under five different situations:
 \{$\delta\zeta_{1}=0$, $\delta\zeta_{2}=0$\},
 \{$0.05\zeta$, $0.05\zeta$\},
 \{$-0.05\zeta$, $0.05\zeta$\},
 \{$0.05\zeta$, $-0.05\zeta$\},
and \{$0.1\zeta$, $-0.1\zeta$\}.

As illustrated in Fig.~6, we find that the collective NV spins will be transferred to the target
state with high efficiency when the dispersion satisfies $|\delta\zeta_{1,2}|<\zeta\times5\%$.
However, the  transfer efficiency will decrease when the dispersion becomes larger.
Hence, in order to prepare the target entangled state with high efficiency in this scheme,
the dispersion of the control parameters  should satisfy $|\delta\zeta_{1,2}|<\zeta\times5\%$.

\begin{figure}[t]
\includegraphics[width=9.5cm]{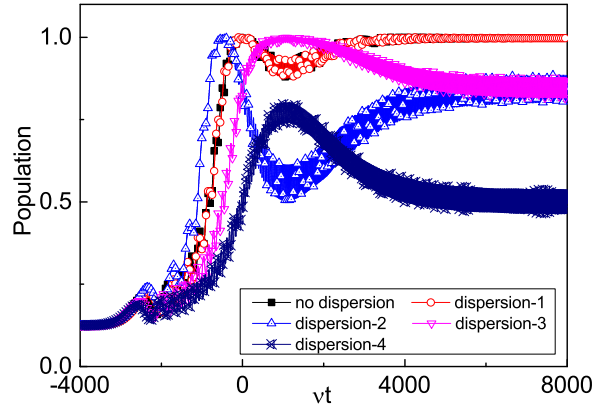}
\caption{\label{fig:wide}(Color online) The dynamical evolution for the population
of the target entangled ground state $|\phi_{(\text{\uppercase\expandafter{\romannumeral1}})}\rangle$
under different dispersion distributions of $\Omega_{1}(t)$ and $\Omega_{2}(t)$,
with the parameters $\lambda=0.1\nu$ and $|\Delta|=0.9\nu$.
The different dispersion distributions are respectively
no dispersion: \{$\delta\zeta_{1}=0$, $\delta\zeta_{2}=0$\},
dispersion-1: \{$0.05\zeta$, $0.05\zeta$\},
dispersion-2: \{$-0.05\zeta$, $0.05\zeta$\},
dispersion-3: \{$0.05\zeta$, $-0.05\zeta$\},
and dispersion-4: \{$0.1\zeta$, $-0.1\zeta$\}.}
\end{figure}

\section{The feasibility of this scheme}

To examine the feasibility of our scheme in realistic experiment, we now
discuss the relevant  experimental parameters.
For realistic conditions,
the frequency for high-$Q$ ($Q\sim10^6$) NAMR is about $\nu/2\pi\sim10$ MHz,
with the number of NV spins $N \sim10$, 
we can obtain the magnetic coupling strength between the NAMR and the NV centers satisfies $\lambda/2\pi\sim 1$ MHz.
The Rabi frequency is about $\Omega _{1,2}/2\pi \sim 1$ MHz and the detuning satisfies $\Delta/2\pi\sim 10$ MHz.
Assuming an environmental temperature $T\sim 10$ mK in a dilution refrigerator, the
thermal phonon number is about $\overline{n}=1/(e^{\hbar\nu/k_{B}T}-1)\sim 20$,
and we can get the effective damping rate of NAMR is about $\gamma_{m}\sim\overline{n}(\frac{\lambda}{\Delta})^{2}\frac{\nu}{Q}<2\pi\times10$ Hz.
Comparing with the effective couplings $|\alpha\beta^{2}\varepsilon|\sim 2\pi\times10$ MHz and  $|\alpha\beta^{2}|\sim 2\pi\times1$ MHz,
we can discard the effect of the NAMR's damping rate in our numerical simulation \cite{RevModPhys.67.249,PhysRevLett.112.213602,PhysRevApplied.4.044003,doi:10.1063/1.1927327, Yang2000Surface}.
Based on these parameters above, the time for transfer the NV spins' ground state adiabatically from separate state to maximal entangled state
will be about $t\sim100$ $\mu s$ in this scheme.
On the other hand, the relaxation time of the NV spin triplet
ranging from milliseconds at room temperature to several seconds at
low temperature has been reported.
In general, the single NV spin decoherence in diamond is mainly caused by the coupling of the surrounding electron or nuclear
spins, such as the electron spins P1 centers, the nuclear spins $^{14}\text{N}$ spins and $^{13}\text{C}$ spins \cite{PhysRevB.83.081201,PhysRevB.85.115303}.
In type-\uppercase\expandafter{\romannumeral1}b diamond samples, the free-induction decay of the NV
center spin in an electron spin bath (P1 centers) can be neglected,
and in high-purity type-\uppercase\expandafter{\romannumeral2}a samples,
the decay time caused by the electron spin bath will exceed one millisecond \cite{PhysRevB.83.081201,PhysRevB.85.115303,PhysRevA.90.032319}.
The coupling to the host $^{14}\text{N}$ nuclear spin($\sim$MHz),
induces the substantial coherent off-resonance errors,
and these errors have been solved experimentally \cite{PhysRevLett.105.200402}.
For NV centers in diamond with natural abundance of $^{13}\text{C}$,
the decoherence will be dominated by the hyperfine interaction with the $^{13}\text{C}$ nuclear spins,
which mainly form the nuclear spin bath \cite{doi:10.1063/1.3666568,Childress2006Coherent,PhysRevLett.106.217205}.
With the development of the dynamical decoupling 
techniques \cite{PhysRevLett.82.2417,PhysRevLett.98.077601,PhysRevLett.98.077602,PhysRevLett.98.100504,Hason2008Coherent,PhysRevLett.101.180403,PhysRevLett.105.230503,PhysRevLett.106.217205,NatBiercuk,NatDuJf,NaturePhysics4, PhysRevLett.101.047601, PhysRevB.82.201201,Zhao2011Atomic},
the dephasing time $T_{2}$  of a single NV center in diamond can be more than $2$ ms \cite{Balasubramanian2009Ultralong,Lange2010Universal,EPL}.
Thus, the coherence time is sufficient for achieving the desired NV spins entangled ground state.

\section{Conclusion}

In summary, we have  proposed an efficient protocol for entangling the NV spins
with the assistance of a high-$Q$ NAMR and dichromatic classical microwave driving fields.
In this protocol,
we can not only acquire the collective LMG type interactions for NV spins ($N\sim 10$),
but also steer the LMG Hamiltonian adiabatically from
the isotropic type to the simple (one-axis twisting) type by tuning
the Rabi frequencies slowly enough to maintain the NV spins in the ground state.
As a result, the collective NV spins will undergo ground state transitions,
which allows us to obtain the adiabatic channels between the initial separate ground state and the final entangled ground state.
In this work, we have made the analytical discussions and numerical simulations on three different types of adiabatic processes
for cases \uppercase\expandafter{\romannumeral1}, \uppercase\expandafter{\romannumeral2}, and \uppercase\expandafter{\romannumeral3}.
We can acquire the GHZ-type maximally entangled NV spin ground state in case \uppercase\expandafter{\romannumeral1},
and the W-type ground states in cases \uppercase\expandafter{\romannumeral2} and \uppercase\expandafter{\romannumeral3}
under realistic conditions.

\section*{Acknowledgments}

This work is supported by the NSFC under Grants No. 11774285,
No. 11474227, and  the Fundamental Research Funds for the
Central Universities. Part of the simulations are coded
in PYTHON using the QUTIP library\cite{Johansson2012QuTiP, Johansson2013QuTiP}.

%

\end{document}